\documentclass[
final,
nomarks
]{dmtcs-episciences}

\usepackage[utf8]{inputenc}
\usepackage{subfigure}
\usepackage{graphicx}
\usepackage{amsmath}
\usepackage{amssymb}  
\usepackage{amsthm}  
\usepackage{xspace}
\usepackage{hyperref}
\usepackage{cleveref}
\usepackage{enumitem}
\usepackage{complexity}
\usepackage{mathrsfs}
\usepackage{doi} 

\newtheorem{theorem}{Theorem}
\newtheorem{lemma}[theorem]{Lemma}
\newtheorem{proposition}[theorem]{Proposition}
\newtheorem{corollary}[theorem]{Corollary}
\newtheorem{remark}[theorem]{Remark}
\newtheorem{problem}[theorem]{Problem}
\newtheorem{question}[theorem]{Question}
\theoremstyle{definition}
\newtheorem{definition}[theorem]{Definition}
\newtheorem{example}[theorem]{Example}

\def\ca#1{{\cal#1}}

\usepackage{tikz}
\tikzstyle{every picture} = [>=latex]
\usetikzlibrary{calc,arrows,decorations.markings,decorations.pathreplacing,quotes,fit,shapes,positioning,backgrounds}

\usepackage[numbers,square]{natbib}

\author[P.\ Hlin\v{e}n\'y and J.~Jedelsk\'y]{Petr Hlin\v{e}n\'y \and Jan Jedelsk\'y}
\title[$k$-Planar and Fan-Crossing Drawings and Transductions]{$k$-Planar and Fan-Crossing Drawings and Transductions of Embeddable Graphs}%
\affiliation{Masaryk University, Brno, Czech Republic}
\keywords{{planar graph; surface embedding; $k$-planar graph; fan-crossing drawing; transduction}}
\begin{document}

\publicationdata{vol. 28:4, SOFSEM 2026}{2026}{6}
	{10.46298/dmtcs.17710}{2026-03-13; 2026-03-13; 2026-06-30}{2026-08-04}

\maketitle
\begin{abstract}\medskip
We introduce, for every surface $\Sigma$, 
a two-way connection between definability of a graph class $\ca C$ 
by FO transductions (first-order logical transformations) of the graphs embeddable in $\Sigma$ 
and a certain variant of fan-crossing drawings of the graphs from $\ca C$ in $\Sigma$.
If $\ca C$ is additionally of bounded maximum degree,
then the restriction on drawings of the graphs from $\ca C$ in $\Sigma$ 
is simply to have a bounded number of crossings per edge
(such as being $k$-planar for fixed~$k$ if $\Sigma$ is the plane).
For graph classes, this connection allows us to derive non-transducibility
results from the nonexistence of the said drawings and, conversely, from
the nonexistence of a transduction to derive nonexistence of the said drawings.
One example of such reasoning is as follows;
since the class of 3D-grids is not transducible from the class of planar graphs,
we can conclude that the class of 3D-grids is not $k$-planar for any fixed~$k$.
On the other hand, the fact that the class of 3D-grids is not $k$-planar for any fixed~$k$
is known also via other means, and this conversely implies that the class of 3D-grids 
is not transducible from the class of planar graphs.
We hope that this connection will help to draw a path to a possible proof
that not all toroidal graphs are transducible from planar graphs.

The result is based on a recent characterization of weakly sparse 
FO transductions of classes of bounded expansion by 
[Gajarsk\'y, G{\l}adkowski, Jedelsk\'y, Pilipczuk and Toru\'nczyk,
{\href{https://arxiv.org/abs/2505.15655}{arXiv:2505.15655}}].
\end{abstract}

\section{Logic, Interpretations, and Drawings of Graphs}\label{sec:intro}

A simple 1-dimensional \emph{first-order (FO) interpretation} of graphs
is given by a binary FO formula $\xi$ over graphs.
For a graph $G$ the result of the interpretation is
the simple graph $\xi(G)$ on the same vertex set and the edge relation
determined by $uv\in E(\xi(G))$ $\iff$ $G\models\xi(u,v)$,
where $G\models\xi(u,v)$ means that the formula $\xi$ is true for the vertex
pair $u,v$ of the graph~$G$.
The formula $\xi$ is assumed symmetric in its parameters.
For example, $\xi(u,v)\equiv\neg edge(u,v)$ interprets the complement of a
graph, and {$\xi(u,v)\equiv edge(u,v)\vee$ $\exists x.\big[u\not=v\wedge
edge(u,x)\wedge edge(x,v)\big]$} interprets the square (second power) of a graph.

In a non-copying \emph{FO transduction} $\tau$, one can additionally
assign arbitrary vertex colors before applying the
$\xi$-interpretation, and then take
an induced subgraph of the result (in particular, $\tau(G)$ is actually a
hereditary set of graphs -- unlike $\xi(G)$).
These notions are naturally extended to graph classes; here $\tau(\ca C)$
denotes the class of all graphs obtained by applying $\tau$ to all (colored)
graphs from a class~$\ca C$.
All transductions in this paper are first-order.
See \Cref{sec:defs} for more detailed and formal definitions
of both notions.

\begin{example}\label{example:transd-subdiv-clique}
	Let $\xi(x, y):= \exists z . C_1(z) \land E(x, z) \land E(z, y)$.
	Consider the non-copying transduction $\tau$ defined by $\xi$. 
	Let $n \in \mathbb{N}$ be arbitrary, and let $G_n$ be
	a 1-subdivision clique $K_n$. Let $H$ be any graph on $n$
	vertices. The edges of $H$ corresponds to a subset $X$ of
	the subdivision vertices of $G$.
	Let $G^+$ be the colored graph obtained from $G$
	by coloring vertices from $X$ by color 1 (and the remaining
	vertices by any other color).
	Then, $\xi(G^+)$ contains $H$ as an induced subgraph.
	Therefore, $\tau(G_n)$ contains all graphs on $\le n$ vertices.
	See \Cref{fig:transd} for an illustration of this example.
\end{example}

\begin{figure}
    \centering
\begin{tikzpicture}[
    scale=0.88,
    vblack/.style={circle, draw=black, fill=black, inner sep=0pt, minimum size=3.4pt},
    vblue/.style={circle, draw=black, fill=blue!70, inner sep=0pt, minimum size=3.4pt},
    vred/.style={circle, draw=black, fill=red!70, inner sep=0pt, minimum size=3.4pt},
    eblack/.style={black, thin},
    ethick/.style={black, thick},
    arrow/.style={->, >=stealth, very thick, gray, shorten <=4pt, shorten >=4pt}
]
    \begin{scope}[shift={(0,0)}]
        \foreach \k in {1,...,5} { \coordinate (v\k) at ({90-(\k-1)*72}:1.2); }
        \foreach \m/\n in {1/2, 1/3, 1/4, 1/5, 2/3, 2/4, 2/5, 3/4, 3/5, 4/5} {
            \coordinate (s\m\n) at ($(v\m)!0.5!(v\n)$);
            \draw[eblack] (v\m) -- (s\m\n) -- (v\n);
            \node[vblack] at (s\m\n) {};
        }
        \foreach \k in {1,...,5} { \node[vblack] at (v\k) {}; }
    \end{scope}
    \draw[arrow] (1.3, 0) -- (3.2, 0);              
    \begin{scope}[shift={(4.5, 0)}]
        \foreach \k in {1,...,5} { \coordinate (v\k) at ({90-(\k-1)*72}:1.2); }
        \foreach \m/\n in {1/2, 1/3, 1/4, 1/5, 2/3, 2/4, 2/5, 3/4, 3/5, 4/5} {
            \coordinate (s\m\n) at ($(v\m)!0.5!(v\n)$);
            \draw[eblack] (v\m) -- (s\m\n) -- (v\n);
        }
        \foreach \m/\n in {1/2, 2/3, 3/4, 4/5, 1/5} { \node[vred] at (s\m\n) {}; }
        \foreach \m/\n in {1/3, 1/4, 2/4, 2/5, 3/5} { \node[vblack] at (s\m\n) {}; }
        \foreach \k in {1,...,5} { \node[vblue] at (v\k) {}; }
    \end{scope}
    \draw[arrow] (5.8, 0) -- (7.7, 0);              
    \begin{scope}[shift={(9.0, 0)}]
        \foreach \k in {1,...,5} { \coordinate (v\k) at ({90-(\k-1)*72}:1.2); }
        \draw[ethick, bend left=22] (v1) to (v2);
        \draw[ethick, bend left=22] (v2) to (v3);
        \draw[ethick, bend left=22] (v3) to (v4);
        \draw[ethick, bend left=22] (v4) to (v5);
        \draw[ethick, bend left=22] (v5) to (v1);
        
        \foreach \m/\n in {1/2, 2/3, 3/4, 4/5, 1/5} { \node[vred] at ($(v\m)!0.5!(v\n)$) {}; }
        \foreach \m/\n in {1/3, 1/4, 2/4, 2/5, 3/5} { \node[vblack] at ($(v\m)!0.5!(v\n)$) {}; }
        \foreach \k in {1,...,5} { \node[vblue] at (v\k) {}; }
    \end{scope}
    \draw[arrow] (10.3, 0) -- (12.2, 0);              
    \begin{scope}[shift={(13.5, 0)}]
        \foreach \k in {1,...,5} { \coordinate (v\k) at ({90-(\k-1)*72}:1.2); }
        \draw[ethick] (v1) -- (v2) -- (v3) -- (v4) -- (v5) -- cycle;
        \foreach \k in {1,...,5} { \node[vblue] at (v\k) {}; }
    \end{scope}
\end{tikzpicture}
    \caption{Illustration depicting how the transduction of \Cref{example:transd-subdiv-clique}
	creates $C_5$ from subdivided $K_5$.
	The leftmost graph is the subdivided $K_5$. To its right, there is one of its
	colorings. Then, there is the graph obtained by the $\xi$-interpretation
	from the colored subdivided clique, where red vertices have color 1 (that is,
	the color used in the formula $\xi$).
	The rightmost graph is an induced subgraph of the interpreted graph.}
    \label{fig:transd}
\end{figure}

We say that a graph class $\ca D$ is \emph{transducible} from a class 
$\ca C$ if there is a transduction $\tau$ such that $\ca D\subseteq\tau(\ca C)$.
We interpret this relation as that $\ca D$ is not ``logically richer''
than~$\ca C$ (although $\ca D$ may look combinatorially a lot more complicated
than~$\ca C$).
Alternatively, the aim is that if we understand and can algorithmically handle
first-order expressible problems on the class $\ca C$, we would be able to do so with~$\ca D$.

We have recently seen a surge of interest in FO transductions of graph
classes, in an effort to develop a model-theoretic structure theory for graphs.
We refer to a recent survey by Pilipczuk \cite{DBLP:journals/corr/abs-2501.04166} for
a broad introduction to this theory, and we only briefly remark that
transductions appear to be the right notion of an ``abstract embedding''
of a graph class in another class for this theory, and that
non-transducibility of ``too complex structures'' is often the right
``limit of tractability'' for the model-checking problem.
We defer further discussion on this topic to \Cref{sec:defs}. 
As such, the transduction hierarchy -- the quasi-order on graph classes given by
the transducibility relation -- is of particular interest.

We remark that, in the case of monadic second-order transductions
(defined similarly, using monadic second-order
logic instead of first-order logic),
there is a long-standing conjecture~\cite{DBLP:journals/corr/abs-1004-4777}
that all graph classes are
MSO-transduction-equivalent to one of the following classes:
single-vertex graph, edgeless graphs, forests of bounded height,
paths, trees, and grids. The only remaining open question in
the MSO-trans\-duction hierarchy asks
whether classes of unbounded clique-width (that is, classes not
MSO-transducible from trees) MSO-transduce the class of all graphs.
This question is also known as the strong Seese conjecture, and
positive answer to this question implies the Seese conjecture
\cite{DBLP:journals/apal/Seese91} saying that classes of unbounded
clique-width have an undecidable MSO theory.

Unlike the MSO transduction hierarchy,
the FO transduction hierarchy is more complex. 
We refer a reader to
\cite[Figure~2]{DBLP:journals/lmcs/BraunfeldNMS25} by Braunfeld,
N\v{e}set\v{r}il, Ossona de Mendez, and Siebertz
for a Hasse diagram of a fragment of the FO transduction hierarchy.
Our article is, in part, motivated by the following open question:
\begin{question}\label{question:hierarchy-genus}
	Is there an integer $t$ such that the class of graphs embeddable on surfaces
	of genus $t+1$ is transducible from those on surfaces of genus $t$?
\end{question}

Transductions are also related to a number of open problems in algorithmic model
theory. For example, Dreier~et~al.~\cite{DBLP:conf/stoc/DreierMT24} proved that
the FO model checking problem is \textsf{AW$[*]$}-hard
(see \cite{FlumGrohe06} for the meaning of \textsf{AW$[*]$}-hard)
even when restricted to any
hereditary class which transduces the class of all graphs, thus resolving one
direction of a conjecture asserting that, a hereditary class $\ca C$
does not transduce the class of all graphs if and only if
the FO model checking problem admits an \textsf{FPT}-time algorithm
when restricted to $\ca C$.

Similarly to the previous conjecture about tractability of the FO model checking problem
and the aforementioned related Seese conjecture about decidability of an MSO theory,
a handful of other logics sitting between the FO and MSO logic (such as so-called FO+conn logic and other)
are similarly conjectured to allow efficiently solvable model checking
on graph classes that do not transduce ``too complex'' classes
(where the exact meaning of ``too complex'' depends on the considered logic).

We hope that a better understanding of the distinction between the classes
transducible from surface-em\-beddable graphs and the classes non-transducible from
them would allow us to find a logic between FO and MSO that
corresponds to surface-embeddable graphs in the following
sense: all the problems expressible in this logic can be solved on
surface-embeddable graphs (and thus likely
also on graph classes transducible from surface-embeddable
graphs), but some of these problems become hard as soon as one considers any
``reasonable'' graph class non-transducible from surface-embeddable graphs.
In order to find such a logic, we must first understand
what characterizes classes transducible from surface-embeddable graphs.
\Cref{question:hierarchy-genus} seems to be the simplest
natural open question in this line of research.

Let $\mathbb{P}$ be a property of graph classes. The model-theoretic structure
theory is, among other things, interested in \emph{structurally-$\mathbb{P}$}
graph classes, that is, the graph classes which are transducible from some graph 
class having the property $\mathbb{P}$. The notion of structurally-$\mathbb{P}$
lets us lift results and structural properties of sparse classes to dense setting.
For instance, structurally bounded tree-depth is bounded shrub-depth
\cite{DBLP:conf/mfcs/GanianHNOMR12}, and structurally bounded tree-width
is stable bounded clique-width \cite{DBLP:conf/soda/NesetrilMPRS21}.
More recently, classes of structurally bounded expansion
\cite{DBLP:journals/tocl/GajarskyKNMPST20,DBLP:journals/lmcs/Dreier23,DBLP:journals/corr/abs-2601-14906} and
structurally nowhere dense \cite{DBLP:conf/stoc/DreierMS23}
have been extensively researched.
Using this language, we thus study weakly sparse
structurally surface-embeddable graph classes.

\medskip

While transducibility is relatively easy to establish in particular
cases of interest (essentially, guess a coloring and the formula), the opposite
(non-transducible) is usually much harder to prove, and existing results are
scarce.

Typically, to prove that a class $\ca D$ is not transducible from a class
$\ca C$, one finds a suitable property of $\ca C$ which is preserved under
transductions and is not satisfied by $\ca D$.
The rather few published examples of properties preserved under transductions
include, e.g.; monadic stability~\cite{SHELAHstabil}, 
bounded clique-width~\cite{DBLP:journals/mst/CourcelleMR00},
each value of shrub-depth~\cite{DBLP:journals/lmcs/GanianHNOM19},
near-uniformness~\cite{DBLP:journals/tocl/GajarskyHOLR20},
bounded twin-width~\cite{DBLP:journals/jacm/BonnetKTW22}
and bounded flip-width~\cite{DBLP:conf/focs/Torunczyk23}.
However, none of these properties is of much help in approaching such a basic and intriguing
question as what characterizes transductions of the \emph{class of planar
graphs}, and more generally, of the class of graphs embeddable in a fixed surface.
So far, to our knowledge, no useful direct connection (of non transducibility) 
to graph topological properties and graph drawings has been published.

Recently, there has been notable progress in two directions
which are related to our effort to understand transductions
of planar and embeddable graphs:
\begin{itemize}[itemsep=0pt,topsep=3pt]
    \item Two groups of researchers \cite{DBLP:conf/lics/GajarskyPP25,DBLP:conf/lics/HlinenyJ25}
independently exploited different properties related to the product structure (of
planar graphs) to prove, among other results, that the class of 3D-grids is
not transducible from planar graphs.
By a \emph{3D-grid} we mean the cartesian product of three paths.
    \item Most importantly for our task, Gajarsk\'y, G{\l}adkowski,
Jedelsk\'y, Pilipczuk and Toru\'nczyk~\cite{DBLP:journals/corr/abs-2505.15655} 
have given a new characterization of transducibility in sparse classes, stated in
\Cref{thm:transducong} below, which resolved several further open questions,
such as that the graphs of tree-width $t+1$ are not transducible from the
graphs of tree-width~$t$ for any~$t$.
On the other hand, classes of bounded clique-width (resp.,~linear clique-width)
are transducible from the class of tree-orders (resp.,~half-graphs).
\end{itemize}

We refer to \Cref{sec:defs} for the definitions of the technical terms used
in the following theorem. We note that planar and surface-embeddable graphs 
are always weakly sparse and of bounded expansion.

\begin{theorem}[Gajarsk\'y et al.~\cite{DBLP:journals/corr/abs-2505.15655}]\label{thm:transducong}
Let $\ca C$ be a graph class of bounded expansion and denote by
$\ca C^\bullet$ the class obtained by adding a universal vertex to
every graph of~$\ca C$.
If $\ca D$ is a graph class transducible from~$\ca C$ such that $\ca D$ is
weakly sparse,
then there exists a $k\in\mathbb N$ such that~$\ca D$ is contained in the
class of congestion-$k$ depth-$k$ minors of the class $\ca C^\bullet$.
\end{theorem}

Whereas Gajarsk\'y et al.~\cite{DBLP:journals/corr/abs-2505.15655} use the \Cref{thm:transducong} primarily in connection with properties of weak colorings in the
concerned graph classes, we give a different, topological, perspective of it.
Our aim is to provide tools from the graph drawing area for proving 
non-transducibility results.

\medskip
We consider only the traditional drawings of graphs in surfaces
(orientable or not) in which no edge passes
through another vertex, no three edges meet in the same point except their
common end, and there are finitely many intersection points
(\emph{crossings} or common end vertices) of pairs of edges.
Tangential intersections of edges, i.e.\ those that can be removed 
by local perturbations of the involved edges, are generally allowed,
but they can be safely avoided in our applications.
We do \emph{not} assume our drawings to be simple.
We often, for simplicity, identify a drawing $D$ of a graph with
the graph represented by $D$ (which allows us to speak about the vertex and edge sets of~$D$, for example).
For a drawing $D$ of a graph, the \emph{crossing graph} $C$ of $D$ has the edges of $D$ as
its vertices, and two edges are adjacent in $C$ if and only if they cross in~$D$.

A drawing of a graph $G$ in a surface $\Sigma$ is a \emph{$k$-crossing} drawing
(e.g., \emph{$k$-planar} if $\Sigma$ is the plane)
if every edge has at most $k$ crossings.
A \emph{fan} in $G$ is any subset of edges incident to the same vertex, called the center.
A drawing is \emph{fan-crossing} if every crossed edge is only crossed by edges of one fan.
If $G'$ is a subdivision of a graph $G$, then an \emph{extended fan} is 
a subgraph of $G'$ formed by the subdivided edges of some fan in~$G$.

\paragraph*{Overview of the results.}
Our main result, \Cref{thm:kfanchar}, is based on a variant of fan-crossing
drawings\footnote{
This generalization somewhat resembles the $k$-fan-bundle-planar drawings of
\cite{DBLP:journals/tcs/AngeliniBKKS18}, but the core difference is
that our ``bundles'' are allowed to branch and cross in multiple sections.
See also \Cref{sec:proofs}.
}
which is analogous to the generalization from $1$-planar to $k$-planar graphs.

\begin{definition}\label{def:clustfan}\rm
A drawing $D$ of a graph $G$ in a surface $\Sigma$ is called 
\emph{$k$-fold $\ell$-clustered fan-crossing} if
there is a drawing $D'$ obtained from $D$ by subdividing each edge with at
most $k-1$ new vertices, such that every connected component of the crossing graph of~$D'$ 
is a subset of the union of the edge sets of at most~$\ell$ extended fans of~$D'$.
Equivalently, there is an assignment of every
crossed edge $e'$ of $D'$ to a fan $F$ of $D$ such that $e'$ is an edge of
the extended fan $F'$ of $F$, and the number of fans altogether assigned 
to the edges of every nontrivial connected component of the crossing graph of~$D'$
is at most~$\ell$.

We furthermore say that a $k$-fold $\ell$-clustered fan-crossing drawing
$D$ as above is \emph{monotone}%
\footnote{Please note that the short conference version 
\cite{DBLP:conf/sofsem/HlinenyJ26} of this paper mistakenly left out the condition
of the $k$-fold $k$-clustered fan-crossing drawings being monotone,
but this condition is implicitly used also there in the proofs.}
if the subdivision $D'$ and the assignment
from the crossed edges of $D'$ to fans of $D$ can be chosen to satisfy also
the following condition:
for every edge $e=uv\in E(D)$, and denoting by $P$ the $u$--$v$ path in $D'$ stemming from $e$,
there is a vertex $w\in V(P)$ (possibly $w=u$ or~$w=v$) such that each crossed edge of the subpath
of $P$ from $w$ to $u$ (resp., from $w$ to~$v$) is assigned to the fan of $D$ centered at $u$ (resp., at~$v$).
\end{definition}
See \Cref{fig:clustfan} for an illustration of a $2$-fold $2$-clustered fan-crossing drawing.
We remark that $1$-fold $\ell$-clustered fan-crossing drawings are not
directly comparable to ordinary fan-crossing drawings which, in general,
do not fulfil the $\ell$-clustered condition for any~$\ell$.
However, a similar `clustering' property is naturally satisfied in $1$-planar 
drawings and in suitable $(k-1)$-subdivisions of $k$-planar drawings.

\begin{figure}[t] \centering
(a)
\begin{tikzpicture}[scale=1.6]\small
\tikzstyle{every node}=[draw, color=black, thick, shape=circle, inner sep=1pt, fill=black]
\node (a) at (-0.6,0) {}; \node (i) at (0.6,0) {};
\node (c) at (-1.6,1) {}; \node (g) at (0,1.1) {};
\node (d) at (-0.6,2) {}; \node (f) at (0.8,2) {};
\node (e) at (0.3,2.2) {};
\draw (c)--(a)--(i) (a)-- +(-1,1.5) (a)-- +(-0.5,2) (a)--(d);
\draw (c)-- +(0.2,0.7) (c)-- +(2.5,-0.8) (c)-- +(2.8,-0.4) (c)--(f);
\draw (i)--(g) (i)--(e) (i)--(f) (i)-- +(0.5,1);
\draw (c)--(g)--(e) (d)--(f)--(e)--(d);
\tikzstyle{every node}=[draw, color=black, thick, shape=circle, inner sep=0.8pt, fill=white]
\node (p) at (-0.2,0.55) {}; \node (q) at (-0.2,0.8) {};
\node (r) at (-0.2,1.58) {}; \node (s) at (0.44,1.2) {};
\begin{scope}[on background layer]
\tikzstyle{every path}=[draw, color=blue!25!white, ultra thick]
\draw (a)-- +(-1,1.5) (a)-- +(-0.5,2) (a)--(d);
\draw (c)-- +(0.2,0.7) (c)--(p) (c)--(q) (c)--(r) (c)--(g);
\tikzstyle{every path}=[draw, color=green!35!white, ultra thick]
\draw (g)--(e)--(s) (d)--(f)--(r);
\tikzstyle{every path}=[draw, color=red!35!white, ultra thick]
\draw (p)-- +(1.1,-0.35) (q)-- +(1.4,-0.2);
\draw (i)--(g) (i)--(s) (i)--(f) (i)-- +(0.5,1);
\end{scope}
\end{tikzpicture}
\qquad\qquad\qquad(b)
\begin{tikzpicture}[scale=1.25]\small
\tikzstyle{every node}=[draw, color=black, thick, shape=circle, inner sep=1pt, fill=black]
\node (a) at (-0.6,0) {}; \node (i) at (0.6,0) {};
\node (b) at (-1.2,0.3) {}; \node (h) at (1.2,0.3) {};
\node (c) at (-1.6,1) {}; \node (g) at (1.6,1) {};
\node (d) at (-1.2,2) {}; \node (f) at (1.2,2) {};
\node (e) at (0,2.5) {};
\draw (a) to[bend right] (c) (b) to[bend right] (d) (c) to[bend right] (e) ;
\draw (d) to[bend right] (f) (e) to[bend right] (g) (f) to[bend right] (h) (g) to[bend right] (i)  ;
\draw (a)--(b)--(c)--(d)--(e)--(f)--(g)--(h)--(i) ;
\node[draw=none,fill=none] at (0,0) {{\boldmath\Large$\ldots$}};
\end{tikzpicture}
    \caption{(a) An example of a monotone $2$-fold $2$-clustered fan-crossing drawing~$D$.
	The four subdivision vertices (i.e., the set $V(D')\setminus V(D)$ from
	\Cref{def:clustfan}) are hollow, and the three components of the crossing
	graph of the subdivided drawing~$D'$ are emphasized by shade colors red, green and blue.
	(b) An ordinary fan-crossing drawing of a large $m$-vertex graph that is
	not $1$-fold $\ell$-clustered for any $\ell < \frac{m}{2}$,
	because no subdivisions are allowed and the crossed edges induce
	$\lfloor\frac m2\rfloor$ edge-disjoint fans.}
    \label{fig:clustfan}
\end{figure}

\begin{theorem}\label{thm:kfanchar}
Let $\Sigma$ be a surface, $\ca C$ be the class of graphs
embeddable in $\Sigma$ and $\ca D$ be a weakly sparse graph class.
Then $\ca D$ is transducible from~$\ca C$, if and only if
there exists $k\in\mathbb N$ such that every graph of~$\ca D$ has 
a monotone $k$-fold $k$-clustered fan-crossing drawing in $\Sigma$
after deleting at most $k$ of~its~vertices.
\end{theorem}

Note that any $k$-fold $k$-clustered fan-crossing drawing of a graph $G\in\ca D$ is
trivially a $(\Delta(G)\cdot k^2)$-crossing drawing in the same surface,
and that deleting some $k$ vertices from $G$ is in this context equivalent to
deleting the at most $\Delta(G)\cdot k$ edges incident to them.
Therefore, we also have the following corollary of \Cref{thm:kfanchar} interesting by itself:

\begin{corollary}\label{cor:kplanchar}
Let $\Sigma$ be a surface, $\ca C$ be the class of graphs
embeddable in $\Sigma$ and $\ca D$ be
a graph class of bounded maximum degree.
Then $\ca D$ is transducible from~$\ca C$, if and only if
there exists $k'\in\mathbb N$ such that every graph of~$\ca D$
has a $k'$-crossing drawing in $\Sigma$ after deleting at most $k'$ of its edges.
\end{corollary}

One can also get a simpler and a bit stronger formulation of these results
if an additional property is satisfied by the class~$\ca D$.
\begin{remark}\label{rem:duplic}
If the `target' class $\ca D$ has the \emph{duplication property},
that is, with each $H\in\ca D$ there is $H'\in\ca D$ such that $H'$ contains
two disjoint copies of $H$ as an induced subgraph,
then the part `after deleting at most $k$ of its vertices/edges' can be 
safely removed from both \Cref{thm:kfanchar} and \Cref{cor:kplanchar}.
\end{remark}
Note that the duplication property is slightly weaker than being closed under disjoint unions,
and it is satisfied, e.g., by the classes of grids and of 3D grids.

\section{Interpretations, Transductions, and Congested Shallow Minors}\label{sec:defs}

We consider only finite simple graphs (forbidding parallel edges
is natural and necessary when dealing with FO transductions
since we view edges as symmetric binary relation; and loops can
be simulated using colors if needed).
When discussing formulas, we implicitly assume the language of colored
undirected graphs with equality. That is, we have the following
atomic formulas: `$x=y$' expressing that the variables $x$ and $y$
refer to the same vertex, `$edge(x, y)$' expressing that $x$ and $y$
are adjacent, and for each color $i$, the atomic formula `$C_i(x)$'
expressing that vertex $x$ has color $i$.
Boolean combinations ($\phi \land \psi$, $\lnot \phi$, $\ldots$) of
formulas are also formulas, and both universal ($\forall x . \phi$)
and existential ($\exists x . \phi$) quantifications (over the graph vertices) 
of formulas are also formulas. An occurrence of a variable $x$ is \emph{bound}
in a formula $\phi$ if it is present within a subformula $\forall x . \psi$
or $\exists x . \psi$ of $\phi$. Otherwise,
the occurrence is \emph{free}. A \emph{free variable} is a variable
with free occurrence. A \emph{sentence} is a formula without free variables.
Given a graph $G$ and a sentence $\phi$, we write $G \models \phi$ to denote
that $G$ is a model of $\phi$, where the domain is the vertex
set $V(G)$ of $G$, and the colors and the edge relation are
interpreted according to the colors and adjacency in $G$.
An FO formula $\xi$ is \emph{binary} if it has only two free variables,
say $x$~and~$y$, and we denote this by writing $\xi(x, y)$.
More generality, we write $\psi(x_1, \ldots, x_k)$ to denote that
$x_1, \ldots, x_k$ are the free variables of $\psi$. Given a graph $G$
and vertices $v_1, \ldots, v_k$, we write $G \models \psi(v_1, \ldots, v_k)$
to denote that $(G, v_1, \ldots, v_k)$ is a model of $\psi(x_1, \ldots, x_k)$
when we modify the language so that each $x_i, i=1,\ldots,k$ becomes a constant
instead of a variable\footnote{Without loss of generality, we assume that
$x_i$ does not have a bound occurrence. Otherwise, we could simply rename
the bound occurrences.}, and the model assigns $v_i$ to $x_i$.
A binary FO formula $\xi(x, y)$ is \emph{symmetric} if
it holds for all colored graphs $G$ that
$G \models \forall x \forall y . \left(\xi(x, y) \leftrightarrow \xi(y, x)\right)$.
We refer the reader to the book Finite model theory by
Ebbinghaus and Flum \cite{DBLP:books/daglib/0082516} for
a more comprehensive introduction to first-order logic on graphs.

\paragraph*{Interpretations.}
Let $\xi(x, y)$ be a symmetric binary FO formula.
Then, $\xi$ defines a mapping from colored graphs to graphs called
\emph{simple 1-dimensional interpretation}.
We refer reader to the aforementioned book \cite{DBLP:books/daglib/0082516}
for definition of general interpretations, and we only briefly and
informally remark that simple and 1-dimensional means that the
output graph has exactly the
same vertices as the input graph.
We only consider simple 1-dimensional interpretations, so we
omit the words ``simple'' and ``1-dimensional'' in the following text.
Furthermore, we only care about interpretation defining graphs,
so the following (very simplified) definition is sufficient here:

Let $G$ be a colored graph. Then,
the \emph{$\xi$-interpretation} of $G$ is a graph $H$
with the same vertex set $V(H)=V(G)$ and the edge relation
determined by $uv\in E(H)$ $\iff$ $G\models\xi(u,v)$.
We write $\xi(G):=H$ to denote the $\xi$-interpretation of $G$. Interpretations
naturally extend to graph classes. For a class $\ca C$, we simply set
$\xi(\ca C):=\{\xi(G) : G \in \ca C\}$.

Informally speaking, an interpretation ``deterministically embeds''
a graph $H=\xi(G)$ into the graph $G$ using the formula $\xi$.
However, the determinism is often too restrictive. We usually consider a slightly
stronger notion called transductions which ``adds non-determinism to interpretations''.

\paragraph*{Transductions.}
A \emph{non-copying transduction} $\tau$ (determined by a symmetric
binary FO formula~$\xi$)
maps a graph $G$ to a set of graphs $\tau(G)$ such that $H\in\tau(G)$
if and only if, for some vertex-coloring $G^+$ of $G$, the graph $H$ is an induced
subgraph of~$\xi(G^+)$.
Informally speaking, a transduction first enhances the graph by adding
a non-deterministically chosen coloring, then it applies an interpretation,
and finally it takes non-deterministically chosen induced subgraph.
See \Cref{example:transd-subdiv-clique} and \Cref{fig:transd} for
a concrete example of a non-copying transduction.

A $k$-copy of a graph $G$ is a structure $kG$ with two binary
relations and $k$ additional unary relations (colors) -- that is, the $edge$
relation, \emph{same-vertex} relation $SV$, and \emph{$i$-th copy}
relation $C_i$ for $i=1,\ldots,k$.
The domain (vertex set) of $kG$ is $V(kG):= \{1,\ldots,k\} \times V(G)$.
A pair of vertices $(i, u), (j, v) \in V(kG)$ is in the edge relation
$E(kG)$ if $i=j$ and $uv \in E(G)$.
A pair of vertices $(i, u), (j, v) \in V(kG)$ is in the same-vertex
relation $SV(kG)$ if $u=v$.
A vertex $(j, u) \in V(kG)$ is in the $i$-th copy relation $C_i(kG)$ if $i=j$.
If $G$ is a colored graph, then we also define colors of $kG$ so that
a vertex $(i, u) \in V(kG)$ has the same colors as $u$.

A \emph{transduction} is a composition of a $k$-copy operation and
non-copying transduction.

We remark that transductions coincide with
non-copying ones on planar graphs and graphs in surfaces, because copying can
be simulated by adding leaves (degree-1 vertices), using more colors, and changing
the formula used in the transduction -- see
\cite[Fact~8.11.]{DBLP:journals/lmcs/BraunfeldNMS25} for a folklore proof.
For simplicity, we stick to this simplified view of transductions without copying.

A class $\ca C$ is \emph{transducible} from a class $\ca D$ if there is
a transduction $\tau$ such that $\ca C \subseteq \tau(\ca D)$. Alternatively, we
sometimes say that $\ca D$ \emph{transduces} $\ca C$. Classes $\ca C$
and $\ca D$ are said to be \emph{transduction-equivalent} if $\ca C$ is transducible
from $\ca D$ and $\ca D$ is transducible from $\ca C$.
A class is always transducible from itself, and
a composition of two transductions is again a transduction, so the
\emph{transducibility relation} is a quasi-order. This quasi-order is called
\emph{transduction hierarchy}.

\paragraph*{Shallow congested minors.}
Consider graphs $G$ and $H$, a collection $\ca A \subseteq 2^{V(G)}$ of
vertex subsets of $G$, and a function $\alpha: V(H) \to \ca A$ such that,
for each $A\in\ca A$, the induced subgraph $G[A]$ is connected.
Then $\alpha$ is called a \emph{congestion-$c$ depth-$d$ minor model} of
$H$ in $G$ (and the members of $\ca A$ are called \emph{model sets}) if:
(i) every induced subgraph $G[A]$, $A\in\ca A$ is of radius~$\leq d$,
(ii) for every vertex $v\in V(G)$, at most $c$ vertices $w \in V(H)$ are
mapped to sets $\alpha(w)$ containing~$v$,
and (iii) for every edge $AB\in E(H)$ (where $A,B\in\ca A$), the sets $A$
and $B$ \emph{touch} in $G$ -- meaning that $A\cap B\not=\emptyset$ or some
two vertices $a\in A$ and $b\in B$ are adjacent in~$G$.
Note that if we set $c=1$, we get the usual (depth-$d$ or unlimited~$d$)
minor model with disjoint model sets.

A graph $H$ is a \emph{congestion-$c$ depth-$d$ minor} of $G$ if
there exists a congestion-$c$ depth-$d$ minor model of $H$ in $G$.
If $c=1$, we speak simply about a \emph{depth-$d$ minor} of $G$.

When discussing congestion-$c$ depth-$d$ minor models $\alpha$ of planar graphs
and graphs of surfaces,
it is often convenient to assume that that no two vertices of $H$ are mapped
to the same set of $\ca A$. We do so whenever we are able to add dummy vertices
to distinguish the sets. In those cases, we further assume that $V(H)=\ca A$, and
we abuse the notation by saying that $\ca A$ is the congestion-$c$
depth-$d$ minor model.

\paragraph*{Sparsity.} A graph class $\ca D$ is \emph{weakly sparse} if there is
an integer $t$ such that no member of $\ca D$ contains a $K_{t,t}$ subgraph.
A graph class $\ca D$ is of \emph{bounded expansion} if there exists
a function $f$ such that every depth-$d$
minor of a graph from $\ca D$ has average degree at most $f(d)$.
Observe that every class $\ca D$ of bounded expansion is necessarily also weakly
sparse. On the other hand, there are weakly sparse graph classes which do not have
bounded expansion (e.g., 1-subdivisions of cliques).

Notably, the classes of planar graphs and of graphs embeddable in a surface are both weakly
sparse and of bounded expansion (see \cite[Section~5.5]{DBLP:books/daglib/0030491}).
Therefore, the aforementioned \Cref{thm:transducong} applies to graph
classes transducible from planar graphs and from graphs embeddable in any fixed surface.

\section{Proofs of \Cref{thm:kfanchar} and \Cref{cor:kplanchar}}\label{sec:proofs}

We start with proving the forward direction `$\Rightarrow$' of both results.

\begin{lemma}\label{lem:todraw}
Let $G$ be a graph embeddable in a surface $\Sigma$ and $H$ be a congestion-$k$ depth-$k$ minor of~$G$.
Then $H$ has a monotone $k'$-fold $k'$-clustered fan-crossing drawing in $\Sigma$ for some $k'\in \ca O(k)$.
\end{lemma}

\begin{proof}
Let $\ca A$ be a congestion-$k$ depth-$k$ minor model of $H$ in $G$,
and for every $v\in V(H)$, let $T_v\subseteq G$ be a rooted spanning tree of
$G[A_v]$ (where $A_v$ represents~$v$) of depth at most $k$, such as a BFS tree of $G[A_v]$.
Let the root of $T_v$ be~$r_v\in V(G)$ (note that $r_v=r_w$ is possible for
distinct $v,w\in V(H)$).

Fix a $\Sigma$ embedding $D$ of $G$, and pick a collection of small pairwise disjoint disks
$\delta_u$ around each $u\in V(G)$, and $\delta_e$ around an internal
(`middle') point of each $e\in E(G)$, such that $\delta_u$ intersects $D$ 
only in edges incident to~$u$ and $\delta_e$ intersects $D$ only in~$e$.

By a \emph{branching arc} with a root $x$ in $\Sigma$ we mean a $\Sigma$
embedding of a tree with $x$ as its root.
The first step is to represent $H$ as a subgraph of the intersection 
graph of a collection of branching arcs $\ca R=\{\varrho_v:v\in V(H)\}$ in $\Sigma$
such that the root of $\varrho_v$ lies in~$\delta_{r_v}$ and the root is disjoint from other
arcs, and that no three branching arcs meet in the same point.
Moreover, for any $v,w\in V(H)$, the branching arcs $\varrho_v$ and
$\varrho_w$ are allowed to intersect only in the disks $\delta_u$ 
for~$u\in A_v\cap A_w$, or in the disk $\delta_e$ if disjoint $A_v,A_w$ 
touch via an edge~$e$~in~$G$.

\begin{figure} \centering
\begin{tikzpicture}[scale=2]\small
\tikzstyle{every node}=[draw, color=black, thick, shape=circle, inner sep=1pt, fill=black]
\tikzstyle{every path}=[draw,black, thin]
\foreach \x in {1,2,...,5} \foreach \y in {1,2,3}  \node (v\x\y) at (\x,\y) {};
\begin{scope}[on background layer]
\foreach \x in {1,2,...,5} \foreach \y in {1,2,3}
	\node[draw=none,fill=black!11!white,inner sep=7pt] at (\x,\y) {};
\foreach \x in {1.5,2.5,3.5,4.5} \foreach \y in {1,2,3}
	\node[draw=none,fill=black!11!white,inner sep=3pt] at (\x,\y) {};
\foreach \x in {1,2,...,5} \foreach \y in {1.5,2.5}
	\node[draw=none,fill=black!11!white,inner sep=3pt] at (\x,\y) {};
\foreach \x / \y / \r in {1/2/8, 1.5/1/6, 3/2/8, 3.5/3/6, 4/1/8, 4/1.5/6, 5/1.5/6, 5/2/8}
	\node[draw=none,fill=black!25!white,inner sep=\r pt] at (\x,\y) {};
\end{scope}
\draw (1,1)--(5,1) (1,2)--(5,2) (1,3)--(5,3) (1,1)--(1,3) (2,1)--(2,3) (3,1)--(3,3) (4,1)--(4,3) (5,1)--(5,3);
\tikzstyle{every path}=[draw,blue, dashed,semithick, rounded corners=2pt]
\draw (0.7,2) to[out=280,in=180] (3,1.7) to[out=0,in=260] (5.3,2) to[out=100,in=0] (3,2.3)  to[out=180,in=80] cycle;
\draw (1,0.7) to[out=10,in=270] (1.3,2) to[out=90,in=350] (1,3.3) to[out=190,in=90] (0.66,2)  to[out=270,in=170] cycle;
\draw (1.7,1) to[out=280,in=180] (3,0.7) to[out=0,in=260] (4.3,1) to[out=100,in=315] (3.45,1.4) to[out=135,in=350] (3,3.3) to[out=190,in=45] (2.55,1.4) to[out=225,in=80] cycle;
\draw (3.7,1) to[out=280,in=180] (4.5,0.7) to[out=0,in=260] (5.3,1) to[out=100,in=0] (4.5,1.3) to[out=180,in=80] cycle;
\draw (3.7,3) to[out=280,in=135] (4.6,2.6) to[out=315,in=170] (5,1.7) to[out=10,in=270] (5.3,3) to[out=105,in=345] (5,3.3) to[out=180,in=80] cycle;
\tikzstyle{every node}=[draw, color=red, thick, shape=circle, inner sep=1.3pt, fill=red]
\tikzstyle{every path}=[draw,red, thick, rounded corners=4pt]
\draw[red!70!yellow] (3.08,0.9)--(4.1,0.9)--(4.1,1.5)--(4,1.55) (3.09,0.9)--(3.09,2.93)--(3.5,2.9)--(3.55,3) (3.09,0.9) node {} --(1.5,0.9)--(1.45,1);
\draw (0.9,3)--(0.9,2) node {} --(0.9,0.93)--(1.5,0.93)--(1.55,1);
\draw[red!85!blue] (4.1,1.9) node[inner sep=0.7pt, fill=black!25!white] {} --(4.1,1.5)--(4,1.45)
	(4.1,1.9)--(2.93,1.9) node {} --(0.9,1.9)--(0.85,1.95)
	(4.1,1.9)--(4.93,1.9) node[inner sep=0.7pt, fill=black!25!white] {} --(5.1,1.9)--(5.15,2) (4.93,1.9)--(4.93,1.5)--(5.13,1.47);
\draw (5.1,0.92) node {} --(4.05,0.92)--(3.97,0.97) (5.1,0.93)--(5.1,1.5)--(5,1.55);
\draw (5.09,2.92) node {} --(5.08,1.5)--(5,1.43) (5.08,2.92)--(3.5,2.92)--(3.45,3);
\end{tikzpicture}
    \caption{An illustration of the proof of \Cref{lem:todraw}.
	The five blue bags (dashed lines in the picture) show the five model sets of a congestion-$2$ depth-$2$ minor model
	of a $5$-vertex graph $H$ in the depicted graph~$G$ (which is a $3\times5$ square grid pictured in black).
	The chosen small disks $\delta_u$ and $\delta_e$ at the vertices and edges of~$G$ are shaded gray.
	The set $\ca R$ of branching arcs representing the vertices of $H$ (so, one per each blue bag) is drawn with thick red lines, each with its highlighted root,
	and the shaded disks in which intersections between the arcs of $\ca R$ occur are emphasized with darker~gray color.
}
    \label{fig:branchingarc}
\end{figure}

Getting the sought drawing of $\ca R$ in $\Sigma$ is nearly straightforward.
See \Cref{fig:branchingarc}.

We draw each $\varrho_v$ of $\ca R$ closely along the (sub)embedding 
of $T_v$ within $D$, but avoiding $\delta_e$ for $e\in E(T_v)$, 
such that for every $vw\in E(H)$;
\begin{itemize}\parskip-2pt
\item if $A_v$ and $A_w$ share a vertex~$u\in A_v\cap A_w$ in the minor model in $G$,
then the arcs $\varrho_v$ and $\varrho_w$ intersect (not necessarily cross) inside $\delta_u$,
and
\item if disjoint $A_v,A_w$ only touch via an edge~$e\in E(G)$, then new
branches of arcs are added to $\varrho_v$ and $\varrho_w$ along the drawing of $e$
such that they intersect inside~$\delta_e$.
\end{itemize}
This drawing can clearly avoid any crossings of arcs outside of the
disks $\delta_u$ and~$\delta_e$.
Note that we only require $H$ to be a subgraph of the intersection graph of
$\ca R$, and so superfluous intersections between the arcs do not pose a problem.
Finally, since multiple intersections between pairs of arcs are allowed, 
this drawing satisfies all desired properties.

Secondly, as $H$ is a subgraph of the intersection graph of $\ca R$,
for every edge $f=vw\in E(H)$ we have a simple arc
$\alpha_{f}\subseteq\varrho_v\cup\varrho_w$ 
between the roots of $\varrho_v$ and $\varrho_w$.
We arbitrarily pick a point in common $b_f\in\alpha_f\cap\varrho_v\cap\varrho_w$.
Then, in a tiny neighborhood of every branching arc $\varrho_v\in\ca R$ we
draw an embedding of a star with the center vertex $v$ in the root of $\varrho_v$ 
and the rays closely following each $\alpha_{f}$ for~$f\ni v$ from $v$ to the point~$b_f$
(the number of rays thus equals the degree of $v$ in~$H$).
This is simultaneously possible for all vertices $v$ of $H$ by our choices of $\alpha_f$ and $b_f$
-- in particular, since the chosen `meeting points' $b_f$ are pairwise
distinct by the property of no three branching arcs meeting in the same point --
and the union of these stars is a drawing $D_1$ of the graph~$H$.
(We note in passing that the obtained drawing easily avoids tangential
intersections between edges, and so has only proper crossings.)

So, in $D_1$, every edge $f=vw\in E(H)$ follows closely a path $P\subseteq G$
such that $P\subseteq T_v\cup T_w\,(+e)$, where $e$ is a possible edge
connecting $A_v$ and $A_w$ if they are disjoint.
Hence, the length of $P$ in $G$ is at most $2k+1$, and we subdivide $f$ of $D_1$ once along
each edge of $P-e$ and twice (before and after~$\delta_e$) at~$e$ if it is present.
After doing so for all $f\in E(H)$, we get a drawing~$D_2$~subdividing~$D_1$.

By the construction, for every connected component $M\subseteq E(D_2)$ of 
the crossing graph of $D_2$ with more than one edge, all crossings of the edges
of $M$ lie either in one of the disks $\delta_u$ for some $u\in V(G)$, 
or in one of the disks $\delta_e$ for some $e\in E(G)$.
In the latter case of $\delta_e$ let $u$ be either end of~$e$.
So, for every edge $f\in E(H)$ such that its subdivision $f'$ in $D_2$
is a path whose edges intersect $M$,
$\>f'$ enters the disk ($\delta_u$ or $\delta_e$) along 
a branching arc $\varrho_v\in\ca R$ where $v\in V(H)$ and $u\in A_v$,
and hence $f$ belongs to a fan of $H$ centered at~$v$.
Since $\ca A$ is a congestion-$k$ model, there are at most $k$ such arcs of
$\ca R$ involved, and so all edges of $M$ in~$D_2$ are contained in the union 
of at most $k$ extended fans in~$D_2$.

In other words, $D_2$ is an $\ca O(k)$-fold $\ca O(k)$-clustered fan-crossing drawing of~$H$,
which is also trivially monotone by our construction.
\end{proof}

The forward direction of \Cref{thm:kfanchar} is now finished by \Cref{thm:transducong} and \Cref{lem:todraw}.

\begin{proof}[of \boldmath`$\Rightarrow$' of \Cref{thm:kfanchar}]
Let $H'\in\ca D$ be a graph. By \Cref{thm:transducong}, there is $G\in\ca C^\bullet$, and a congestion-$k$ depth-$k$ minor model of
$H'$ in $G$. Let $u\in V(G)$ be a universal vertex such that $G-u$ is embeddable in~$\Sigma$. 
Let $U\subseteq V(H')$ be the set of at most $k$ vertices
of $H'$ represented in this model by sets containing~$u$.
Then $H:=H'-U$ is a congestion-$k$ depth-$k$ minor of the graph $G-u$ which is embeddable in~$\Sigma$, 
and we conclude with the drawing and bound $k'$ by \Cref{lem:todraw}.
\end{proof}

The forward direction of \Cref{cor:kplanchar} then follows easily as mentioned in \Cref{sec:intro}.

\begin{proof}[of \boldmath`$\Rightarrow$' of \Cref{cor:kplanchar}]
For each $H'\in\ca D$,
the previous proof yields a $\ca O(k)$-fold $\ca O(k)$-clustered fan-crossing drawing of $H=H'-U$ in $\Sigma$,
which is at the same time a $\ca O(\Delta(H')\cdot k^2)$-crossing drawing of~$H$ in $\Sigma$.
Since $|U|\leq k$, by adding the vertices of $U$ as isolated vertices into this drawing,
we obtain a $\ca O(\Delta(H')\cdot k^2)$-crossing drawing of~$H'$ after deleting 
$\ca O(\Delta(H')\cdot k)$ of its edges incident to~$U$.
\end{proof}

The converse `$\Leftarrow$' directions of both \Cref{thm:kfanchar} and \Cref{cor:kplanchar}
are stated (as parts (b) and (a), respectively) in the next \Cref{lem:drawingtofo}.
Although part (a) of \Cref{lem:drawingtofo} is actually implied by part (b), 
we prefer to prove the parts separately since (a) is much easier 
and it accessibly illustrates the necessary concepts.

\begin{lemma}\label{lem:drawingtofo}
Let $k\in\mathbb N$,  $H$ be a graph and $X\subseteq V(H)$ where $|X|\leq k$ be a vertex set such that $H-X$ has
\begin{itemize}\parskip-2pt
\item[a)] a $k$-crossing drawing in~$\Sigma$, or
\item[b)] a monotone $k$-fold $k$-clustered fan-crossing drawing in~$\Sigma$.
\end{itemize}
Then there is an FO formula $\xi_k(x,y)$ depending only on~$k$, 
and a colored graph $G$ on vertex set $V(G) \supseteq V(H)$ embedded in $\Sigma$
such that~$H$ equals $\xi_k(G)[V(H)]$ (the subgraph of $\xi_k(G)$ induced on~$V(H)$).
\end{lemma}

\begin{figure} \centering
\begin{tikzpicture}[xscale=1.4,yscale=2]\small
\tikzstyle{every node}=[draw, color=black, thick, shape=circle, inner sep=1pt, fill=black]
\node (a) at (-0.8,0) {}; \node (i) at (0.8,0) {};
\node (c) at (-1.6,1.2) {}; \node (g) at (1.6,1.2) {};
\node (e) at (0,2) {};
\draw (a)--(c)--(e)--(g)--(i)--(a) to[bend left=11] (e) (e) to[bend left=11] (i);
\draw (i) to[bend left=11] (c) (c) to[bend left=11] (g) (g) to[bend left=11] (a) ;
\end{tikzpicture}
\qquad\raise 6ex\hbox{\LARGE$\leadsto$}\qquad
\begin{tikzpicture}[xscale=1.4,yscale=2]\small
\tikzstyle{every node}=[draw, color=black, thick, shape=circle, inner sep=1pt, fill=black]
\node (a) at (-0.8,0) {}; \node (i) at (0.8,0) {};
\node (c) at (-1.6,1.2) {}; \node (g) at (1.6,1.2) {};
\node (e) at (0,2) {};
\draw (a)--(c)--(e)--(g)--(i)--(a) to[bend left=11] (e) (e) to[bend left=11] (i);
\draw (i) to[bend left=11] (c) (c) to[bend left=11] (g) (g) to[bend left=11] (a) ;
\tikzstyle{every node}=[draw, color=black, thick, shape=circle, inner sep=1.2pt, fill=red!90!black]
\draw node at (0,0.25) {} node at (0.67,0.56) {} node at (0.37,1.37) {};
\draw node at (-0.67,0.56) {} node at (-0.37,1.37) {};
\tikzstyle{every node}=[draw, color=black, thick, shape=circle, inner sep=0.9pt, fill=white]
\draw node at (-0.77,0.15) {} node at (0.77,0.15) {};
\draw node at (-0.1,1.85) {} node at (0.1,1.85) {};
\draw node at (-1.2,1.27) {} node at (1.2,1.27) {};
\draw node at (-1.2,0.9) {} node at (1.2,0.9) {};
\draw node at (-0.55,0.08) {} node at (0.55,0.08) {};
\tikzstyle{every node}=[draw, color=black, thick, shape=circle, inner sep=1.2pt, fill=blue!90!black]
\draw node at (-0.73,0.33) {} node at (0.73,0.33) {} ;
\draw node at (-0.6,0.83) {} node at (-0.5,1.1) {} ;
\draw node at (0.6,0.83) {} node at (0.5,1.1) {} ;
\draw node at (-0.25,1.6) {} node at (0.25,1.6) {} ;
\draw node at (0.2,0.33) {} node at (-0.25,0.17) {} ;
\tikzstyle{every node}=[draw, color=black, thick, shape=circle, inner sep=1.2pt, fill=green!70!black]
\draw node at (-0.73,1.33) {} node at (0.73,1.33) {} ;
\draw node at (-0.1,1.38) {} node at (0.1,1.38) {} ;
\draw node at (-0.89,0.7) {} node at (0.89,0.7) {} ;
\draw node at (-0.45,0.45) {} node at (0.45,0.45) {} ;
\draw node at (-0.2,0.33) {} node at (0.25,0.17) {} ;
\end{tikzpicture}
    \caption{An illustration of the proof of \Cref{lem:drawingtofo}(a) (here with $X=\emptyset$);
	turning a $2$-planar drawing $D$ into a planar colored drawing $D''$.
	Color $b_0$ is red, and $b_1$ and $b_2$ are green and blue.
}
    \label{fig:clustkplan}
\end{figure}

\begin{proof}
(a)
We let $D$ be a $k$-crossing drawing of the graph $H-X$ in~$\Sigma$, 
and make it a $\Sigma$ embedding $D'$ by turning each crossing of $D$ into a new vertex.
Let $D''$ subdivide every edge of $D'$ incident to a former crossing twice
and $G''$ be the graph of $D''$.
Let $X=\{x_1,\ldots,x_m\}$,~$m\leq k$, and
$G$ be formed from $G''$ by adding $x_1,\ldots,x_m$ as isolated~vertices.
Note that $V(H)\subseteq V(G)$
and every edge of $H-X$ has a corresponding
(i.e., formed by the said subdivisions) 
path in $G$ of length at most~$3(k+1)$.
See \Cref{fig:clustkplan} for an illustration in the plane.

We now color the vertices of $G$ by sets of colors
(which can easily be transformed into singleton colors at the end), as follows.
For simplicity let us say that we \emph{give a color} to a vertex~$v$,
which means that the resulting color of $v$ will be the collection 
of all colors given to $v$ throughout the construction of coloring.

We introduce colors $c_i$ and $c'_i$ for $1\leq i\leq m$ and three more
colors $b_0$, $b_1$ and~$b_2$ in~$G$.
Color $c_i$ is given to $x_i$ and $c'_i$ to every neighbor of $x_i$ in~$H$.
Color $b_0$ is given to every vertex $v$ of $G$ coming from a crossing in $D$.
If such $v$ comes from a crossing of two edges in $D$, we arbitrarily
order these edges as $e_1,e_2$, and for each $i\in\{1,2\}$, we give color $b_i$ to
the two neighbors of $v$ in $G$ which belong to the former edge~$e_i$~of~$D$.

The formula $\xi_k(x,y)$ is constructed as a disjunction of the following possibilities:
\begin{itemize}\parskip-2pt
\item For some $1\leq i\leq m\leq k$, the vertex $x$ contains color $c_i$ and $y$ color $c'_i$ or vice versa, or
\item $xy$ is an edge of the graph~$G$, or
\item there exists a path $P=(x,z_1,\ldots,z_p,y)$ in $G$ of length at most $3(k+1)$,
such that for every $1<i<p$ the vertex $z_i$ contains color $b_1$ or
$b_2$, or $z_i$ contains color $b_0$ and the color sets of its neighbors $z_{i-1}$ and $z_{i+1}$
both contain the same color among $b_1$ or $b_2$.
\end{itemize}
The said properties are routinely expressible in FO logic, giving the desired FO formula $\xi_k(x,y)$ over $G$.

Note that $\xi_k$ depends on the parameter $k$ in two ways;
in the first point where a disjunction of $m\leq k$ possibilities 
for color pairs $c_i$ and $c_i'$ occurs,
and in the third point in which we use a separate existential 
quantifier for every internal vertex of the path~$P$.
On the other hand, the construction of $\xi_k(x,y)$ depends neither on the actual graph~$G$, nor on the surface~$\Sigma$.

It is also immediate that, when restricted to the ground set $V(H)$,
$\xi_k(x,y)$ captures precisely the edges $xy$ of~$H$.
Indeed, all edges of $H$ with one or both ends in $X$ are captured by the first point,
all edges of $H-X$ which are not crossed in~$D$ are captured by the second point
(note that no new edges with both ends in $V(H)$ have been created),
and every edge crossed in $D$ has been turned into a colored path in $G$ 
accepted by the third point of $\xi_k(x,y)$.
The only slightly less trivial task is to check that every colored path in $G$
accepted by $\xi_k(x,y)$ comes from an actual edge of $D$, which follows from
our coloring in $G$ and the way it is checked by $\xi_k(x,y)$.

\medskip(b)
Let $D$ be a monotone $k$-fold $k$-clustered fan-crossing drawing of~$H-X$ in~$\Sigma$,
and let $D'$ be the subdivision of $D$ as assumed by \Cref{def:clustfan}.
First of all, by local perturbations, we may assume that there are no
tangential intersections between edges in $D'$, only proper crossings.
Let $D_1$ further subdivide every crossed edge of $D'$ incident to a vertex 
$v\in V(H)$ right next to $v$ (for technical reasons, we do not want
original vertices of $H$ to be incident to crossed edges).
Let $H_1$ be the graph drawn by~$D_1$
and $M_i\subseteq E(H_1)$, $i=1,\ldots,a$, be the nontrivial ($|M_i|>1$)
connected components of the crossing graph of~$D_1$.
Observe that there is a natural one-to-one correspondence between these components
and the nontrivial connected components of the crossing graph of~$D'$:
each $M_i$ corresponds to a component $M_i'$ of the crossing graph of~$D'$
such that $M_i\setminus M_i'$ consists of edges formed by subdivisions
of the edges of $M_i'\setminus M_i$ when forming $D_1$ from~$D'$.

For every $i\in\{1,\ldots,a\}$ we introduce the following notation.
By \Cref{def:clustfan}, there are altogether at most $k$ fans 
$F_i^1,\ldots,F_i^k\subseteq E(H-X)$ assigned to all edges of~$M_i$.
Let $w_i^j$ denote the center vertex of the fan~$F_i^j$,
and $\bar F_i^j\subseteq H_1$ the extended fan which is a subdivision of~$F_i^j$.
We denote by $R_i^j\subseteq V(H_1)$ the subset of the vertices
incident to $M_i\cap E(\bar F_i^j)$ which are reachable from $w_i^j$ in $\bar F_i^j-M_i$
(informally, $R_i^j$ can be seen as the ``entry points'' of $\bar F_i^j$ to~$M_i$),
and by $T_i^j\subseteq V(H_1)$ the subset of the vertices
incident to $M_i\cap E(\bar F_i^j)$ which are not~in~$R_i^j$.

Let $D_2$ denote the drawing -- actually an embedding in $\Sigma$, obtained
from $D_1$ by turning every crossing into a new vertex.
Next, independently for each $i\in\{1,\ldots,a\}$, we contract all new vertices
coming from the crossings in~$M_i$ into one vertex~$m_i$ 
(along an arbitrarily chosen spanning tree) while removing loops.
Hence, we get an induced star $S_i$ centered in~$m_i$ and all leaves being
from $V(H_1)\setminus V(H)$, and we denote by $S_i'$ the graph obtained by
subdividing each ray of $S_i$ with one new vertex.
Doing this in $D_2$ simultaneously for all $i=1,\ldots,a$,
we altogether get a drawing $D_3$ which again is an embedding in $\Sigma$
since contractions of non-loop edges preserve embeddability.
See \Cref{fig:clustcontr} for details.

\begin{figure}[tb] \centering
\begin{tikzpicture}[xscale=1.4, yscale=2.4]\small
\draw[dotted,rounded corners=3mm, fill=black!10!white] (-0.8,-0.1) rectangle (1.5, 1.1) {};
\node[draw=none] at (1.1,0.3) {$M_i$} ;
\tikzstyle{every node}=[draw, color=black, thick, shape=circle, inner sep=1pt, fill=black]
\node[label=left:$w_i^2$, label=below:$\quad\ca F_i^2$] (w1) at (-2,0) {};
\node[label=left:$w_i^1$, label=above:$\quad\ca F_i^1$] (w2) at (-2,1) {};
\node[label=right:$w_i^1$, label=left:$\ca F_i^3\quad$] (w3) at (0.7,-0.5) {};
\draw (w1)-- +(4,0) (w1)-- +(4,0.6) (w1)-- +(4,1.2);
\draw (w2)-- +(4,0) (w2)-- +(4,-0.6) (w2)-- +(4,-1.2);
\draw (w3)-- +(-1.8,1.1) (w3)-- +(-0.95,2) (w3)-- +(0.1,2);
\draw[densely dotted] (-0.9,0.17) ellipse (0.08 and 0.26);
\draw[densely dotted] (-0.9,0.84) ellipse (0.08 and 0.26);
\draw[densely dotted] (0.45,-0.2) ellipse (0.4 and 0.05);
\tikzstyle{every node}=[draw, color=black, thick, shape=circle, inner sep=0.9pt, fill=white]
\node[label=below:$~R_i^2~$] (a) at (-0.9,0) {}; \node (b) at (-0.9,0.17) {}; \node (c) at (-0.9,0.33) {};
\node (d) at (-0.9,0.48) {}; \node (e) at (-0.9,0.67) {}; \node (f) at (-0.9,0.84) {};
\node[label=above:$~R_i^1~$] (g) at (-0.9,1) {};
\node (h) at (-0.15,1.3) {}; \node (i) at (0.79,1.3) {}; \node (j) at (1.7,1.11) {};
\node (k) at (1.7,1) {}; \node (l) at (1.7,0.56) {}; \node (m) at (1.7,0.44) {};
\node (n) at (1.7,0) {}; \node (o) at (1.7,-0.11) {}; \node (p) at (0.2,-0.2) {};
\node (q) at (0.55,-0.2) {}; \node[label=right:$~R_i^3$] (r) at (0.71,-0.2) {};
\end{tikzpicture}
\quad\raise16ex\hbox{\LARGE$\leadsto$}\quad
\begin{tikzpicture}[xscale=1.4, yscale=2.4]\small
\tikzstyle{every node}=[draw, color=black, very thin, shape=circle, inner sep=1pt, fill=none]
\node[label=left:$w_i^2$, label=below:$\quad\ca F_i^2$] (w1) at (-2,0) {};
\node[label=left:$w_i^1$, label=above:$\quad\ca F_i^1$] (w2) at (-2,1) {};
\node[label=right:$w_i^1$, label=left:$\ca F_i^3\quad$] (w3) at (0.7,-0.5) {};
\tikzstyle{every path}=[draw, dotted, thin]
\draw (w1)-- +(4,0) (w1)-- +(4,0.6) (w1)-- +(4,1.2);
\draw (w2)-- +(4,0) (w2)-- +(4,-0.6) (w2)-- +(4,-1.2);
\draw (w3)-- +(-1.8,1.1) (w3)-- +(-0.95,2) (w3)-- +(0.1,2);
\tikzstyle{every path}=[draw]
\tikzstyle{every node}=[draw, color=black, thick, shape=circle, inner sep=1.1pt, fill=white]
\node (a) at (-0.9,0) {}; \node (b) at (-0.9,0.17) {}; \node (c) at (-0.9,0.33) {};
\node (d) at (-0.9,0.48) {}; \node (e) at (-0.9,0.67) {}; \node (f) at (-0.9,0.84) {};
\node (g) at (-0.9,1) {};
\node (h) at (-0.15,1.3) {}; \node (i) at (0.79,1.3) {}; \node (j) at (1.7,1.11) {};
\node (k) at (1.7,1) {}; \node (l) at (1.7,0.56) {}; \node (m) at (1.7,0.44) {};
\node (n) at (1.7,0) {}; \node (o) at (1.7,-0.11) {}; \node (p) at (0.2,-0.2) {};
\node (q) at (0.55,-0.2) {}; \node (r) at (0.71,-0.2) {};
\node[fill=black, label=above:$~\quad m_i\quad~$] (s) at (0.4,0.5) {};
\draw (a)--(s)--(b) (c)--(s)--(d) (e)--(s)--(f) (g)--(s)--(h) (i)--(s)--(j) ;
\draw (k)--(s)--(l) (m)--(s)--(n) (o)--(s)--(p) (q)--(s)--(r) ;
\node[fill=red!70!white] (aa) at (-0.7,0.08) {}; \node[fill=red!70!white] (b) at (-0.7,0.22) {};
\node[fill=red!70!white] (c) at (-0.7,0.35) {};
\node[fill=green!50!black] (d) at (-0.7,0.48) {}; 
\node[fill=blue!50!white] (e) at (-0.7,0.64) {}; \node[fill=blue!50!white] (f) at (-0.7,0.78) {};
\node[fill=blue!50!white] (g) at (-0.7,0.92) {};
\node[fill=green!50!black] (h) at (-0.05,1.15) {}; \node[fill=green!50!black] (i) at (0.72,1.15) {};
\node[fill=red!80!black] (j) at (1.5,1.02) {}; \node[fill=blue] (k) at (1.5,0.92) {}; 
\node[fill=red!80!black] (l) at (1.5,0.55) {}; \node[fill=blue] (m) at (1.5,0.45) {};
\node[fill=red!80!black] (n) at (1.5,0.08) {}; \node[fill=blue] (o) at (1.5,-0.02) {}; 
\node[fill=green!90!black] (p) at (0.25,0) {};
\node[fill=green!90!black] (q) at (0.5,0) {}; \node[fill=green!90!black] (r) at (0.63,0) {};
\end{tikzpicture}
    \caption{An illustration of the proof of \Cref{lem:drawingtofo}(b)
	(here with $X=\emptyset$).
	(left) Component $M_i$ of the crossing graph of the drawing $D_1$ from the proof,
	with sets $\ca F_i^1$, $\ca F_i^2$, $\ca F_i^3$ of paths subdividing
	original fans centered at $w_i^1$, $w_i^2$,~$w_i^3$.
	(right) The corresponding fragment of the embedding $D_3$, displaying
	the subdivided star $S_i'$ and the colors given by the proof.
	Color $b_0$ is black, $b_0'$ is white, and $b_j,b_j'$, $j=1,2,3$,
	are light and dark (resp.) shades of colors in order blue, red and green.
	The light shades of colors in the picture are given to the vertices of $R_i^j$,
	and the dark ones to those of $T_i^j$.
}
    \label{fig:clustcontr}
\end{figure}

Similarly as in (a), we let $G$ be the graph obtained from the graph
drawing $D_3$ by adding the isolated vertices of $X=\{x_1,\ldots,x_m\}$,~$m\leq k$.
In $G$, we introduce colors $c_i$ and $c'_i$ for $1\leq i\leq m\leq k$,
and colors $b_j$ and $b'_j$ for $0\leq j\leq k$, as follows:
\begin{itemize}\parskip-2pt
\item Color $c_i$ is given to $x_i\in X$ and $c'_i$ to every neighbor of $x_i$ in~$H$.
\item Color $b_0'$ is given to every vertex of $V(H_1)\setminus V(H)$
(these are the degree-$2$ vertices created by subdivisions forming $D'$ and then~$D_1$),
and color $b_0$ is given to all vertices $m_1,\ldots,m_a$ (which have been
created by contractions of the components of the crossing graph of~$D_1$).
\item For every $i\in\{1,\ldots,a\}$ and $j\in\{1,\ldots,k\}$, 
color $b_j$ is given to the vertices of $S_i'$ which subdivide
the edges of $S_i$ from $m_i$ to $R_i^j$,
and color $b_j'$ is given to the vertices of $S_i'$ which subdivide
the edges of $S_i$ from $m_i$ to $T_i^j$.
\end{itemize}

The formula $\xi_k(x,y)$ over the graph $G$ is constructed as a disjunction of the following possibilities:
\begin{itemize}\parskip-2pt
\item For some $1\leq i\leq m$, the vertex $x$ is given color $c_i$ and $y$ color $c'_i$ or
vice versa, or
\item $xy$ is an edge of~$G$, or
\item there exist $0<p\leq k$, a walk $P=(x,z_0,\ldots,z_{4p},y)$ in $G$
(so, $P$ is of length $4p+2\leq4k+2$), and $0\leq q\leq p$,
such that the following conditions hold
\begin{itemize}\parskip-1pt
\item all vertices $z_{4i}$ for $0\leq i\leq p$ contain color $b_0'$,
\item all vertices $z_{4i+2}$ for $0\leq i<p$ contain color $b_0$, and
\item for every $0\leq i<p$ there is $j\in\{1,\ldots,k\}$ such that 
the vertex $z_{4i+1}$ contains color $b_j$ and $z_{4i+3}$ contains color $b_j'$ if~$i<q$,
and $z_{4i+1}$ contains color $b_j'$ and $z_{4i+3}$ contains color $b_j$ if~$i\geq q$.
\end{itemize}
\end{itemize}
Again, as in part (a), the described properties are routinely expressible in FO logic,
in particular using a separate existential quantifier for every internal vertex of the walk~$P$,
and the construction of the FO formula $\xi_k(x,y)$ depends neither on the graph $G$ nor on the surface~$\Sigma$.

It is left to verify that the formula $\xi_k(x,y)$ where $x,y\in V(H)$ defines the edge set of the graph~$H$,
more precisely that $G\models\xi_k(x,y)$ if and only if $xy\in E(H)$.
If $x,y\in V(H)$ are such that $\{x,y\}\cap X\not=\emptyset$, then, since the vertices of $X$ have no edges in~$G$,
we have $G\models\xi_k(x,y)$ if and only if $xy\in E(H)$
by the first point of the definition of~$\xi_k$.

So, from now on, we consider pairs $x,y\in V(H)$ such that $\{x,y\}\cap X=\emptyset$.
Since the only edges of $G$ having both ends in $V(H)$ are the uncrossed edges of~$D$
(all crossed edges of $D$ have been subdivided in our construction),
we get that $G\models\xi_k(x,y)$ if $xy$ is uncrossed in~$D$, which is if and only if $xy\in E(H)\cap E(D_2)$.
Two possibilities remain to be considered; (i) $xy\in E(H)\setminus E(D_2)$, and (ii) $xy\not\in E(H)$.

In the case (i), the edge $xy$ of $H$ is, by our construction, 
subdivided into a path $P'_{xy}$ in the drawing~$D_2$, 
and then contracted into a walk%
\footnote{Note that (although it is impossible in the drawing produced by \Cref{lem:todraw})
\Cref{def:clustfan} admits a situation in which one ray of an extended fan in the drawing $D'$
intersects a component $M_i$ of the crossing graph of $D'$ in more than one edge,
and in such case the contracted vertex $m_i$ in $G$ may be repeated in $P$ more than once.
This does not influence the arguments in our proof.}
$P_{xy}$ in the drawing~$D_3$.
Note that only the vertices $z_{4i+2}$ for $0\leq i<p$ (those given color $b_0$) may be repeated in~$P$.
Since the original drawing $D$ is $k$-fold, at most $k-1$ of the vertices of $P$ may be from
$V(D')\setminus V(D)$ and at most $2$ other from $V(D_1)\setminus V(D')$.
These vertices are given color $b_0'$ in~$G$, and so they are among $z_{4i}$ for $0\leq i\leq p$;
we have $p\leq k-1+2=k+1$ and the length of $P_{xy}$ is at most~$4k+2$.
Then, by our coloring of the vertices of $G$, the walk $P_{xy}$ is accepted
by the third point in the definition of~$\xi_k$, 
where $q$ is determined by the monotonicity property in \Cref{def:clustfan}.
This means that $G\models\xi_k(x,y)$, as needed.

In the remaining case (ii) we have $x,y\in V(H-X)$ such that $xy\not\in E(H)$,
and we want to show that $G\not\models\xi_k(x,y)$.
Let $\bar F_x$ and $\bar F_y$ denote the extended fans of $H_1$ centered at $x$ and $y$, respectively.
By means of contradiction, we already know that $G\models\xi_k(x,y)$
could be true only if there exist a walk $P=(x,z_0,\ldots,z_{4p},y)$ in $G-X$
and $0\leq q\leq p$ with the properties claimed by the definition of $\xi_k$.
Our aim is to prove, in the next paragraph, that $z_{4q}\in V(\bar F_x)\cap V(\bar F_y)$.
Considered in the graph $H_1$ which subdivides~$H$, 
this claim then means that $\bar F_x\cap\bar F_y$ contains
a path of $H_1$ connecting $x$ to $y$ and avoiding $V(H)$ otherwise.
In other words, that~$xy\in E(H)$ which is a contradiction.

By symmetry, it suffices to prove $z_{4q}\in V(\bar F_x)$ which we do inductively for $q'=0,1,\ldots,q$.
Our claim is trivial for $q'=0$ since $z_0$ is a neighbor of $x$ in~$H_1$.
For an arbitrary $0\leq q'<q$, assume that $z_{4q'}\in V(\bar F_x)$ by induction.
Let $i\in\{1,\ldots,a\}$ be such that $z_{4q'+2}=m_i$, and let
$j\in\{1,\ldots,k\}$ be such that $\bar F_i^j=\bar F_x$, 
and so $z_{4q'}\in R_i^j$ by our definitions.
See the illustration in \Cref{fig:clustcontr}.
Then $z_{4q'+1}$ is colored $b_j$ in $G$, and hence $z_{4q'+3}$ is colored $b_j'$ 
by the formula $\xi_k$, which means that $z_{4q'+4}\in T_i^j$ by the definition
of our coloring of~$G$.
The latter fact, in turn, means that $z_{4q'+4}=z_{4(q'+1)}\in V(\bar F_i^j)=V(\bar F_x)$
and our claim holds for~$q'+1$.

This completes the proof.
\end{proof}

\begin{proof}[of \boldmath`$\Leftarrow$' of \Cref{thm:kfanchar} and \Cref{cor:kplanchar}]
If a graph class $\ca D$ satisfies the assumptions of \Cref{cor:kplanchar}
(resp., of \Cref{thm:kfanchar}) for some $k$ and some surface~$\Sigma$,
then every graph $H\in\ca D$ satisfies the conditions of \Cref{lem:drawingtofo}\,a) 
(resp., of \Cref{lem:drawingtofo}\,b)) for the same $k$ and~$\Sigma$.
Since the graph $G$ constructed in the proof of \Cref{lem:drawingtofo}
is embeddable in $\Sigma$, and the constructed FO formula $\xi_k$ does not
depend on the particular graph $G$, we thus get an FO transduction of the class
of all graphs embeddable in $\Sigma$ that contains all graphs of~$\ca D$.
\end{proof}

\section{Potential Applications}

Inspired by the recent interest in studying transducibility
between graph classes, and namely from the class of planar graphs,
we have proved (based on~\cite{DBLP:journals/corr/abs-2505.15655}) an
asymptotic characterization of transducibility from planar graphs 
and from graphs embeddable in any fixed surface
in the weakly sparse world -- \Cref{thm:kfanchar}.

For example, since 3D-grids are of maximum degree~$6$,
from \cite{DBLP:conf/lics/GajarskyPP25,DBLP:conf/lics/HlinenyJ25},
\Cref{cor:kplanchar} and \Cref{rem:duplic} we immediately conclude:
\begin{corollary}\label{cor:3dgridk}
The class of 3D-grids is not $k$-planar for any fixed~$k$,
and not $k$-crossing for any fixed $k$ on any fixed surface.
\end{corollary}

We are not aware of any published elementary proof of \Cref{cor:3dgridk}, though,
this claim, at least in the planar case, can be easily derived from published results:
Dujmovi\'c et al.~\cite{DBLP:journals/siamdm/DujmovicEW17}
proved that every $n$-vertex $k$-planar graph has tree-width 
$\ca O\big(\sqrt{(k+1)n}\big)$
and, on the other hand, Dvo\v{r}\'ak and Wood showed
\cite{DBLP:journals/corr/abs-2208-10074}
that any balanced bipartition of an $n$-vertex 3D grid is crossed
by $\Omega(n^{2/3})$ edges, and so the tree-width is at least $\Omega(n^{2/3})$.
Knowing validity of \Cref{cor:3dgridk} via other means,
we can also use \Cref{cor:kplanchar} to conversely prove that 3D-grids are not
transducible from planar graphs (this is, in addition to
\cite{DBLP:conf/lics/GajarskyPP25,DBLP:conf/lics/HlinenyJ25}, a third way of
proving the statement which was open only two years ago).

Nevertheless, our main desire is to draw a path to a possible solution
of the following specific subproblem of \Cref{question:hierarchy-genus},
in which \emph{toroidal graphs} are the graphs embeddable in the torus.
\begin{problem}[cf.~\cite{DBLP:journals/corr/abs-2501.04166}]\label{prb:toroid}
Is the class of toroidal graphs transducible from that of planar graphs?
\end{problem}
By \Cref{thm:kfanchar}, a positive answer to Problem~\ref{prb:toroid} is equivalent to having
a monotone $k$-fold $k$-clustered fan-crossing drawing (after deleting $\leq k$ of its
vertices) for fixed $k$ and every toroidal graph, which does not seem
likely to us.
An even more presentable connection exists in the case of bounded degrees,
via \Cref{cor:kplanchar} (note that \Cref{rem:duplic} does not apply here):
\begin{problem}\label{prb:torodeg}
For which $d\geq3$ (all of them?) does there exist $k$ such that every
toroidal graph of maximum degree at most $d$ is $k$-planar after deleting 
at most $k$ of its edges?
\end{problem}
\noindent
\Cref{prb:torodeg} makes good sense independently of \Cref{prb:toroid}, even
without allowing the deletion of $\leq k$ edges.

A negative answer to \Cref{prb:torodeg} (for any~$d$) would readily give
a negative answer to \Cref{prb:toroid}.
On the other hand, a positive answer to \Cref{prb:torodeg},
even for any single~$d\geq3$, confirms an affirmative answer to \Cref{prb:toroid}
for all bounded-degree classes of toroidal graphs; see \Cref{pro:anydegs}.
If, at the same time, \Cref{prb:toroid} had a negative answer in general, this would uncover
an interesting structural difference between bounded-degree toroidal graphs and all toroidal graphs.

\begin{proposition}\label{pro:anydegs}
Assume there exist integers $d\geq3$ and $\ell$ such that every
toroidal graph of maximum degree at most $d$ is $\ell$-planar 
after deleting at most $\ell$ of its edges.
Then every class of toroidal graphs of bounded maximum degree
is transducible from the class of planar graphs.
\end{proposition}
\begin{proof}
We build on the following two standard observations:
\begin{enumerate}[label=(\roman*)]\parskip-2pt
\item If $H$ is an $\ell$-planar graph, then $H$ is 
a congestion-$2$ depth-$\ell$ minor of a planar graph~$H_1$.
\item If $H$ is a depth-$\ell_1$ minor of a congestion-$c$ 
depth-$\ell_2$ minor of a graph $H_1$, then $H$ is 
a congestion-$c$ depth-$(2\ell_2+1)\ell_1$ minor of $H_1$.
\end{enumerate}
For (i), we construct the planar graph $H_1$ by introducing one new vertex 
for every crossing of the $\ell$-planar drawing of $H$ and replacing
every edge $e$ of $H$ with a path $Q_e$ whose internal vertices 
are the vertices of the crossings on $e$ in order.
A minor model of $H$ then assigns the internal vertices of every
such path $Q_e$ to the model set of an arbitrary one of the ends of $e$.
For (ii), we simply ``stack'' one minor model on top of the other,
which does not increase the congestion.

\smallskip
Let now $\ca D$ be a class of toroidal graphs of maximum degree $\Delta$, 
and let $G\in\ca D$ be embedded in the torus.
For every vertex $v\in V(G)$, we replace $v$ with a path $P_v$ on $deg_G(v)$ vertices,
and make edges formerly incident to $v$ now incident each to a different vertex of $P_v$
in order given by the rotation of these edges in embedded $G$.
Denoting the resulting graph by $G_1$, we easily get that $G_1$ is
toroidal of maximum degree $\Delta_1=3\leq d$, and $G$ is a depth-$\Delta$ minor of $G_1$.

By the assumption (of \Cref{pro:anydegs}), 
there is an induced subgraph $G_1'\subseteq G_1$
obtained by deleting $\leq\ell$ vertices, such that $G_1'$ has an $\ell$-planar drawing,
and so $G_1'$ is a congestion-$2$ depth-$\ell$ minor of a planar graph $G_2$ by (i).
Moreover, by the definition of a minor model, 
there is an induced subgraph $G'\subseteq G$ obtained by deleting $\leq\ell$ vertices,
such that $G'$ is a depth-$\Delta$ minor of $G_1'$.
Consequently, by~(ii), $G'$ is a congestion-$2$ depth-$(2\ell+1)\Delta$ minor of $G_2$,
and by \Cref{lem:todraw}, $G'$ has a monotone $k'$-fold $k'$-clustered fan-crossing
drawing for some $k'\in \ca O(\ell\Delta)$.
Therefore, in particular,
every $G\in\ca D$ has a $k$-fold $k$-clustered fan-crossing drawing
after deleting at most $k$ of its vertices for $k=\max(\ell,k')$.
By \Cref{thm:kfanchar}, we conclude that $\ca D$ is transducible from the class of planar graphs.
\end{proof}

\section{Final Remarks}

We conclude with some thoughts on future research specifically of interest
to the graph drawing world.

Firstly, \Cref{def:clustfan} is strongly inspired by the classical fan-crossing drawings.
With our current knowledge, we cannot say whether it is a genuine generalization
of fan-crossing drawings.
In particular, \Cref{fig:clustfan} shows that \Cref{def:clustfan} does not include 
all fan-crossing drawings when $k=1$.
We hence conjecture that there exist (probably ``small'') integers $k,\ell$ 
such that every fan-crossing drawing is a $k$-fold $\ell$-clustered fan-crossing drawing.
This conjecture may be easier to verify in the realm of simple drawings than in general.

Second, we propose another generalization of fan-crossing drawings,
similar in spirit to \Cref{def:clustfan}, which is as follows.
Let a drawing $D$ be called \emph{strictly $k$-fold fan-crossing}
if there is a drawing $D'$ obtained from $D$ by subdividing each edge
at most $k-1$ times, such that every edge of $D'$ is only crossed by edges
belonging to one extended fan of~$D'$.
Note that for $k=1$, this definition in the plane exactly describes ordinary fan-crossing 
drawings, but there does not seem to be a direct relation to \Cref{def:clustfan}.
Two interconnected questions are thus natural;
are there nontrivial inclusions between the classes of $k$-fold
$\ell$-clustered fan-crossing graphs and the classes of strictly $k'$-fold
fan-crossing graphs for some parameters $k,\ell,k'$,
and can \Cref{thm:kfanchar} be proved also for strictly $k$-fold
fan-crossing drawings?

Lastly, Hendrey, Karol and Wood \cite{DBLP:journals/corr/abs-2507-22395}
have recently defined $k$-matching-planar drawings as the drawings $D$
such that, for every edge $e$ of $D$, the edges of $D$ crossing $e$ do not
admit a matching of size~$k+1$.
Every strictly $k$-fold fan-crossing drawing in the plane is
$k$-matching-planar, but the converse direction does not seem easy to handle.
A relation of $k$-matching-planar drawings to our \Cref{def:clustfan} is
unclear in either direction, and this may be an interesting topic for
further research in graph drawing.

\acknowledgements
Both authors have been supported by the project 26-21334S of the Czech Science Foundation.


\bibliographystyle{abbrvnat}
\bibliography{Hcw,trcr}

\end{document}